# Density Functional Theory Analysis of Na$_3$AgO: Assessing its Viability as a Sustainable Material for Solar Energy Applications$^\star$

Vipan Kumar[a,e], Shyam Lal Gupta[b], Sumit Kumar[c], Ashwani Kumar[d], Pooja Rana[a,*] and Diwaker[e,**]

[a]*Department of Physics, Sri Sai University, Palampur, 176061, H P , INDIA*
[b]*Exploring Physics for Interdisciplinary Science and Technology (EPIST) Lab, Harish Chandra Research Institute, Prayagraj, Allahabad, 211019, U P , INDIA*
[c]*Department of Physics, Government College, Una, 174303, H P , INDIA*
[d]*School of Basic Sciences, Abhilashi University, Mandi, 175045, H P , INDIA*
[e]*Department of Physics, S. C. V. B. Government College, Palampur, Kangra, 176061, H P , INDIA*

ABSTRACT

This study mainly emphasis's the fascinating features of inverse perovskite's Na$_3$AgO using density functional theory (DFT). Inverse perovskites (IPs) Na$_3$AgO structural features have been examined, and the space group and cubic structure of Pm-3m (221) have been confirmed. The experimental formulation and thermal stability of IPs have been confirmed by the formation energy. Phonon dispersion curves were used to assess dynamic stability. The dynamic stability of the examined IPs and the bonding strength against cubic structure deformation are confirmed by the lack of negative frequencies. The energy gap ($E_g$) or the characteristics of semiconducting behaviour have been predicted by the electronic properties of Na$_3$AgO with a band gap of 1.273 eV. In order to confirm the viability of solar cells, the light-dependent properties have also been identified. Born stability criteria are also used to verify the mechanical stability, and additional elastic characteristics are identified in order to forecast the anisotropy, ductility, strength, and hardness. These anti-perovskites, which possess intriguing characteristics, have the potential to be effective materials for photovoltaic applications, as indicated by the analysed findings.

## 1. Introduction

The growing energy demands of the modern industrialized world have spurred a lot of research into efficient energy harvesting devices inside antiperovskite compositions. Oxide perovskites have garnered substantial attention in the realms of technical advancements, fundamental research, and commercial applications due to their adaptable properties. Nevertheless, despite the extensive investigation, researchers worldwide have failed to consider additional categories of materials that are closely associated with oxide perovskites, particularly antiperovskites (APs,). These APs' varied physical characteristics raise serious scientific concerns. Depending on their chemical makeup, these materials can display a variety of properties, such as magnetic, semiconductor, and superconductive ones. Antiperovskites have the same layout as perovskites with formulation ABX$_3$, but they differ in that the atomic positions of the A and X elements are switched, resulting in an inversion between cations and anions (A$_3$BX). These structural modifications could lead to distinctive physical traits and a wide range of technological applications [1, 2, 3, 4, 5, 6, 7, 8]. Given their numerous applications, it is imperative to gain more knowledge about the physical characteristics of APs. The chemical and physical characteristics of alkali metal-based oxide APs, one of the known antiperovskite materials, are still mostly unknown. These compounds' basic physical and chemical properties are exciting to study because of the distinctive arrangement of their chemical constituents. There aren't many studies on these chemicals in the literature currently under publication, despite their remarkable electrical and bonding properties. Antiperovskites have shown suitable properties for use in thermoelectrics, photovoltaics, energy storage, and superconductivity. Sr$_{(3-x)}$SnO was previously produced using a simple process and was said to possess superconducting characteristics. Additionally, Ca$_{(3-x)}$XO (X = Pb, Sn) was synthesized, exhibiting cubic geometry and higher room temperature thermoelectric power generation characteristics. In order to assess the TE performance of Ca$_3$XO (X = Si, Ge), Mar et al. used an experimental and computational method. These APs' low heat conductivity and better power factor confirmed that their superior performance was possible. A number of theoretical researchers were inspired to simulate oxide APs and forecast their fascinating features for upcoming energy technologies by the scant experimental investigations of these materials [9, 10]. In their analysis of Ba$_3$SiOs vibrational properties and electronic nature, Castro et al. [11] demonstrated how the structural change from cubic to orthorhombic caused an electronic transition from metallic behavior to insulating features. The optical properties and TE features of (Li/Na/K)$_3$OI were disclosed by Liang et al. [12], who also shown that these APs had significant absorption, making them competitive substitutes for perovskite solar cells. The Rb-based antiperovskites were examined by Sharma et al.[13], who found certain physical characteristics that were appropriate for renewable energy systems. The aforementioned studies have inspired us to investigate the specific physical properties of Ag-based APs and alkali metals for photovoltaic applications, such as their dynamic stability, mechanical stability, and optoelectronic qualities. In the current work, we employed Density Functional Theory (DFT) to investigate the effects of sodium (Na) on the multifunctional properties of materials. Our primary objective was to analyze the stability of compounds represented as Na$_3$AgO along with their bandgap energy and photoelectric response. This knowledge aims to enhance the potential applications of these materials in energy harvesting. We conducted a comprehensive analysis of the Na$_3$AgO compounds, focusing on their electronic band structure, density of states, and stability within a cubic framework. Moreover, we explored the optical dielectric function, which facilitated the calculation of other interesting optical characteristics of Na$_3$AgO. We are hopeful that our findings will shed light on the effects of doping on these properties and contribute to the development of more environmentally friendly energy technologies based on Na$_3$AgO.

## 2. Computational details

Our theoretical investigations in this study, which were based on the Density Functional Theory (DFT) idea, were carried out using the WIEN2K package. The Full-Potential Linearised Augmented Plane Wave (FP-LAPW) technique, which is renowned for its remarkable accuracy in resolving the Kohn–Sham equations, is used by WIEN2K [14, 15, 16, 17, 18, 19, 20, 21]. This methodology is very suitable for the analysis of complex structural systems with a variety of chemical and electrical contexts because it takes into account the framework's full capabilities without the need for geometric estimations. To do the structural evaluation, we employed Perdew, Burke, and Ernzerhof's (PBE) Generalised Gradient Approximation (GGA). Because it provides a better representation of exchange-correlation energies than the Local Density Approximation (LDA), especially in systems with notable changes in electronic concentration, GGA-PBE is widely employed. Accurate calculations of lattice parameters and other structural characteristics are guaranteed by this approach. We investigated electrical and optical characteristics using GGA-PBE and the modified Becke-Johnson (mBJ) exchange potential. The mBJ potential is particularly useful for improving semiconductor material bandgap estimates, which are typically underestimated by standard DFT computations like LDA or GGA. The parameter $K_{Max} \times R_{MT}$ was chosen to be 7 in this theoretical framework, where $R_{MT}$ is the Muffin-Tin (MT) sphere radius encircling each and every atom inside the unit cell, and $K_{Max}$ is the absolute value of the biggest k-vector in the plane-wave expansion. This number determines the size of the basis set, making it an essential component of FP-LAPW calculations. For the majority of crystalline materials, a value of 7 is frequently sufficient to achieve convergence and produce reliable findings for attributes. $L_{max} = 10$ was the upper limit of the angular momentum expansion for the wavefunctions inside the Muffin-Tin spheres. At $G_{max} = 12$ (in atomic units), the interstitial region's potential's Fourier expansion came to an end. To ensure the accuracy of the self-consistent field (SCF) calculations, strict convergence constraints were set. While the charge density change was limited to $10^{-5}$ atomic units, the energy difference between two consecutive rounds was kept to less than $10^{-5}$ Ry. For reliable results, these exacting standards are essential, particularly for attributes that are electrical structure-sensitive. The Brillouin zone integration method was tetrahedral, with a mesh consisting of 216 k-points. Dense k-point sampling provides high-resolution information on the electrical properties and ensures overall energy convergence [22]. This approach makes it easier to fully understand the material's electrical structure and how it affects optoelectronic and energy device applications.

## 3. Results and Discussion
### 3.1. Structural properties

An A$_3$BX stoichiometry is produced when the positions of the cation (A-site) and anion (X-site) are switched in antiperovskite structures, often referred to as inverse perovskites (IPs), a class of inorganic materials with a perovskite-like framework (ABX$_3$). Inverse perovskites have the X-site at the center and the A-sites at the edges because of the inverted octahedral arrangement, which sets them apart from conventional perovskites. As shown in Fig. 1, the Na$_3$AgO oxide IPs satisfy the structural requirements, with oxygen occupying the center of the octahedra and the Ag atoms positioned at the corners. Since the oxidation states in the examined IPs are Na$^+_3$Ag$^-$O$^{2-}$, the octahedra can center on the O rather than the B-site, which is common in ordinary perovskites. Pm-3m (221) is the cubic space group that includes Na$_3$AgO oxide IPs. In order to correctly determine lattice parameters and other ground-state properties, this structure was optimized. Fig. 1b displays the parabolas that were produced by optimizing the structure. The parameters of the structure were then predicted by fitting the curves using the Murnaghan equation of state given as.

$$E_{tot}(V) = E_0(V) + \frac{B_0 V}{B_0(B_0 - 1)} \left[ B\left(1 - \frac{V_0}{V}\right) + \left(\frac{V_0}{V}\right)^{B'_0} - 1 \right] \quad (1)$$

Additionally, Table 1. contains the values for the lattice constant a(Å), ground state volume ($V_o$), Bulk Modulus (B), ground state energy ($E_o$) and formation energy $E_f$.

Before employing any compound in technical domains, the substance should also be stable under various operating conditions. Formation energy ($E_f$) helps predict a substance's thermal stability and acts as a thermal indication of breakdown risk. To increase perovskites' performance in technological applications at high temperatures, their chemical composition and arrangement must be optimized to obtain low formation energies. For our IP Na$_3$AgO the

**Table 1**
Optimized structural and other parameters of IP Na$_3$AgO

| IP | Phase | a(Å) | $V_o$(GPa) | B(Gpa) | $E_o$(Ryd) | $E_f$(eV/atom) |
|---|---|---|---|---|---|---|
| Na$_3$AgO | NM | 4.66 | 683.0251 | 25.55 | -11759.809600 | -2.27 |
| Rb$_3$AgO [23] | NM | 5.64 | - | 13.79 | - | -0.19 |

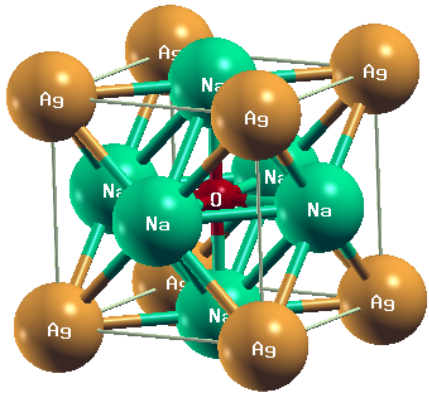

(a)

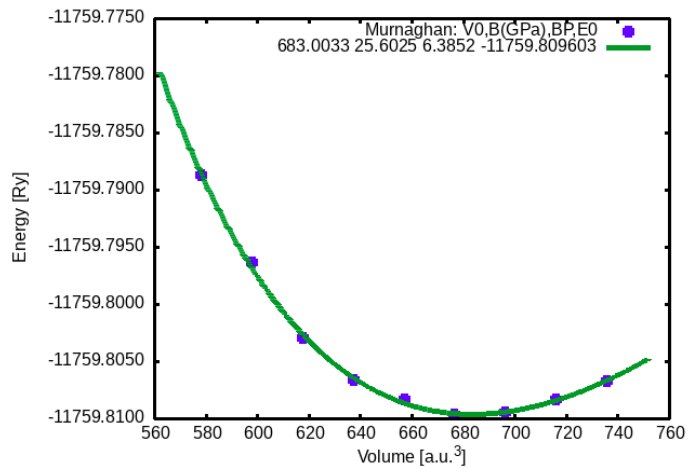

(b)

**Figure 1:** [a] The crystal structures of the cubic IP Na$_3$AgO [b] Energy Vs Volume curve for IP Na$_3$AgO

formation energy is calculated by the following equation given as

$$\Delta E_f = \frac{E_{tot}(Na_3AgO) - (3E_{Na} + E_{Ag} + E_O)}{5} \quad (2)$$

where $\Delta H_f$ $E_{tot}$, $E_{Na}$, $E_{Ag}$ and $E_O$ are the energy of formation, the optimal total energy of the IP Na$_3$AgO, the energy of the Na atom's ground state, the lowest energy state of the Ag atom, and the lowest energy state of the O atom. The investigated IP structural stability is demonstrated by the negative value of its formation energy, which is approximately -2.27 eV/atom. This -ve value of formation energy ascertain the possibility of experimental synthesis of IP Na$_3$AgO and confirms its thermodynamic stability.

## 4. Electronic Properties

The calculated density of states(TDOS/PDOS) for Na$_3$AgO is shown in fig. 3(a-d). It is clear from the figure that the density of states can be divided into two regions. The first region is valence band range from -2.4 eV to 0 eV with a major contribution from Ag-f states while small contribution from O-d states. The 2nd region is conduction band, which ranges from 1.2 eV to 8 eV. In this band, the prominent contribution comes from Na-p states with a noticeable part from Na-s states In optoelectronic and photovoltaic applications the energy band gap and density of states is important electronic characteristics of material under study [24]. The energy band gap fig. 2a of Na$_3$AgO is 1.273 eV, hence it is suitable for visible and infrared applications. Fig 3. shows the DOS and PDOS for Na$_3$AgO which reveals the electronic composition band structure of the material under study. The maximum of VB occur at 100.9037 eV and minimum of VB occur at 5.6582 eV in DOS curve. From the total Density of States plot it is clear that the valence band of the alloy has a high density of state indicating a large number of occupied energy levels. Also, the conduction band of the alloys has a low state density, indicating fewer energy levels available for conduction. At -4.1876 eV, Na$_3$AgO has the highest occupied state in the valence band, and at 2.1653 eV, conduction band has the highest occupied states. The small overlap near the Fermi level forms the band gap, spreading electron density from the valence band to the conduction band. In valance band the main contribution is from Ag-f states and the conduction band consists mainly with Na-2p states near the fermi level, while fermi level is situated at 0 eV

## 5. Mechanical Stability and Properties

Understanding the response of Na$_3$AgO anti-perovskites for mechanical behavior requires an understanding of elasticity in materials under stress. When the energies of the deformed structural configurations are higher than the energies of the first phases, these materials are said to have a mechanically stable cubic structure [25]. The elastic constants (ECs), which are based on measurements, are used to evaluate the stiffness and durability of the material. To examine the ECs without external pressure, tensor and stress values for microscopic strains are required. Instead of taking into account the whole set of elastic constants ($C_{ij}$), just three distinct ones $C_{11}$, $C_{12}$, and $C_{44}$ are taken into account for cubic structural symmetry. The elastic constants for the cubic structure are computed using the IRelast code which is integrated into the WIEN2k software. By satisfying the Born conditions for stability, the expected ECs ensure the mechanical stability of the materials being studied. it is expressed by the following equations as below [26]

$$C_{11} - C_{12} > 0 \quad (3)$$
$$C_{11} > 0 \quad (4)$$
$$C_{11} + 2C_{12} > 0 \quad (5)$$
$$C_{12} < B < C_{11} \quad (6)$$
$$C_{44} > 0 \quad (7)$$
$$(8)$$

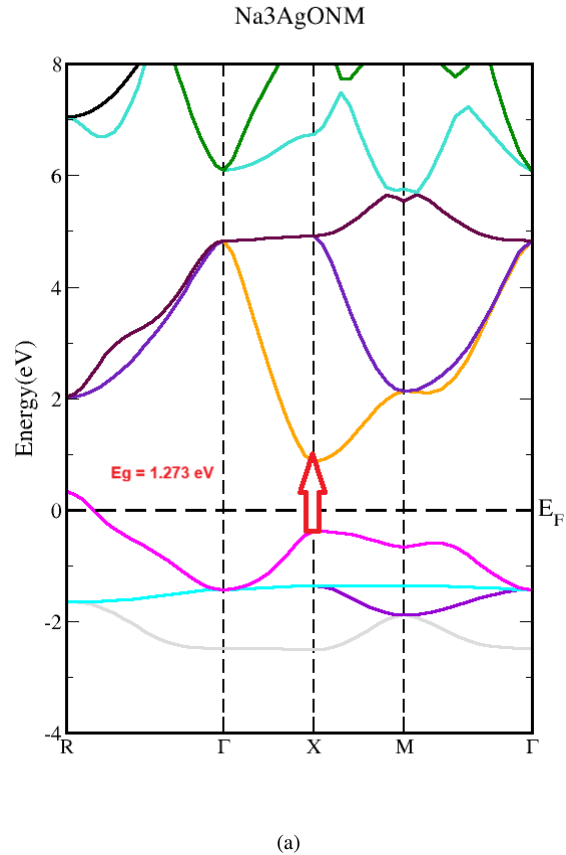

(a)

**Figure 2:** [a] BandGap of cubic IP Na$_3$AgO

Other elastic properties, such as bulk modulus (B), shear modulus (G), Poisson's ratio ($\nu$), and Young's modulus (Y), can be computed using the formulas below.

$$B = \frac{(C_{11} + 2C_{12})}{3} \quad (9)$$
$$Y = \frac{9BG}{3B + G} \quad (10)$$
$$\nu = \frac{3B - 3G}{2(3B + G)} \quad (11)$$
$$G = \frac{G_v + G_R}{2} \quad (12)$$

**Table 2**
List of important mechanical parameters of Na$_3$AgO anti-perovskites

| Parameter | Na$_3$AgO |
|---|---|
| Elastic constant $C_{11}$ (in GPa) | 32.131 |
| Elastic constant $C_{12}$ (in GPa) | 30.320 |
| Elastic constant $C_{44}$ (in GPa) | 19.375 |
| Bulk Modulus B (in GPa) | 30.924 |
| Young's Modulus E (in GPa) | 31.847 |
| Shear Modulus G (in GPa) | 11.987 |
| Pugh's ratio ($\frac{B}{G}$) (in GPa) | 4.386 |
| Anisotropy index A (in GPa) | 23.337 |
| Lame's $1^{st}$ constant $\lambda$ (in GPa) | 26.223 |
| Lame's $2^{nd}$ constant $\mu$ (in GPa) | 7.051 |
| Kleinman parameter $\zeta$ (in GPa) | 1.672 |
| Poisson's Coefficient $\nu$ (in GPa) | 0.328 |
| Transverse Velocity (in m/s) | 1492.752 |
| Longitudnal Velocity (in m/s) | 3569.803 |
| Average Velocity (in m/s) | 1688.444 |
| Debye Temperature $\theta_D$ (in K) | 184.460 |
| Vickers hardness (in GPa) | 0.683 |

$$(13)$$

All important mechanical parameters for Na$_3$AgO anti-perovskites are tabulated in table. 2. A material's resistance to volumetric change under uniform compression is measured by its bulk modulus [27]. When exposed to external compressive loads, a material with a greater B has a better ability to maintain volume. Our computed value of B listed in table. 2. indicates that (B < 40 GPa) Na$_3$AgO anti-perovskites material seems to be flexible in nature [28]. The resistance of a material to shape deformation under shear stress is indicated by its G [29]. The computed value of shear modulus for our material is 11.987 GPa which indicates that under applied stress our material is less resistant to shape deformation. The Young's Modulus Y of a material can be used to determine its stiffness and strength. The computed value of Y for our AP Na$_3$AgO comes out to be 31.847 GPa which indicates that it is less robust material in nature. When assessing the ductility of materials, two crucial metrics are the Poisson's ratio ($\nu$) and Pugh's ratio (B/G) [30]. If a material's ($\nu$) is more than 0.26, it is deemed ductile; if not, it is brittle. The Pugh's ratio (B/G) is calculated by dividing the bulk modulus (B) by the shear modulus (G). If B/G is more than 1.75, it exhibits ductile behavior; if not, it is brittle. The computed value of Poisson's ratio ($\nu$) and Pugh's ratio (B/G) for our AP Na$_3$AgO comes out to be 0.328 and 4.386 respectively. these values indicate that our composition is ductile. The relationship shown below indicates a correlation between the anisotropy factor (A) and the material's ability to form microcracks [31].

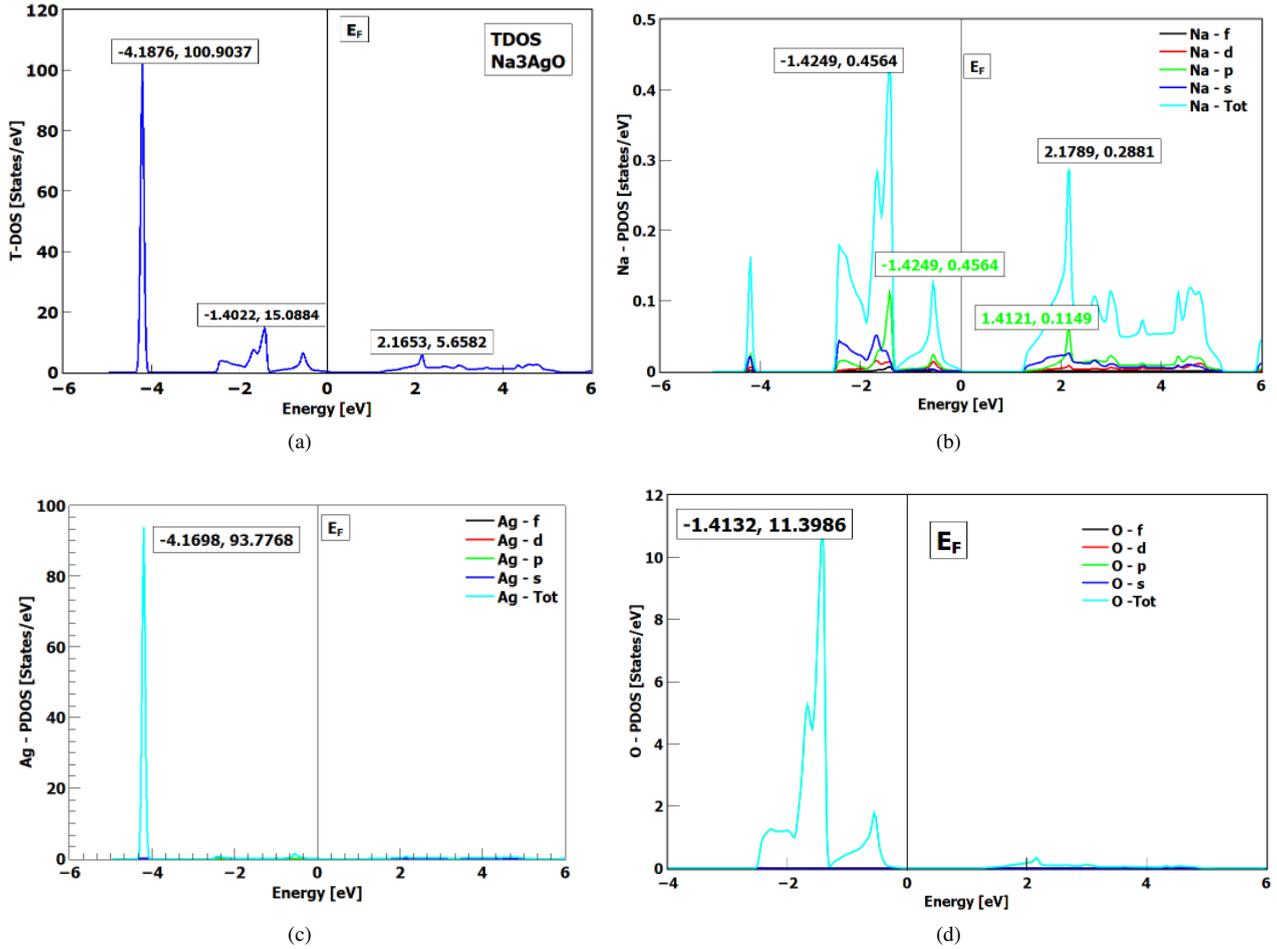

**Figure 3:** TDOS and PDOS of IP Na$_3$AgO

$$A = \frac{2C_{44}}{C_{11} - C_{12}} \qquad (14)$$

Anisotropic materials are defined as those with A $\neq$ 1, and isotropic materials are defined as those with A = 1. While isotropic materials have constant mechanical characteristics in all directions, anisotropic materials exhibit directional dependency in their properties. The materials under investigation which is AP Na$_3$AgO are substantially anisotropic as the value deviates from 1. For each step, the mechanical properties, such as B, G, Y, and $\nu$, are calculated using anisotropic elastic coefficients. The ELATE tool is used to calculate the mechanical properties and anisotropic behavior [32]. ELATE is a specialized tool for researching elastic moduli and viewing anisotropic mechanical properties. The 2D and 3D graphs for the materials Na$_3$AgO displayed in fig. 2. validate anisotropic characteristics. Any deviation from unity denotes asymmetric behavior, whereas A = 1 denotes symmetric behavior. Maximum and lowest elastic modulus values (Y, G, and $\nu$) along with associated anisotropic components (A) are shown in Table 3. Our results show that these elastic moduli's A values differ from 1, indicating that asymmetrical shapes are revealed by these anti-perovskites. The results show how these materials can be used in engineering and other technological domains.

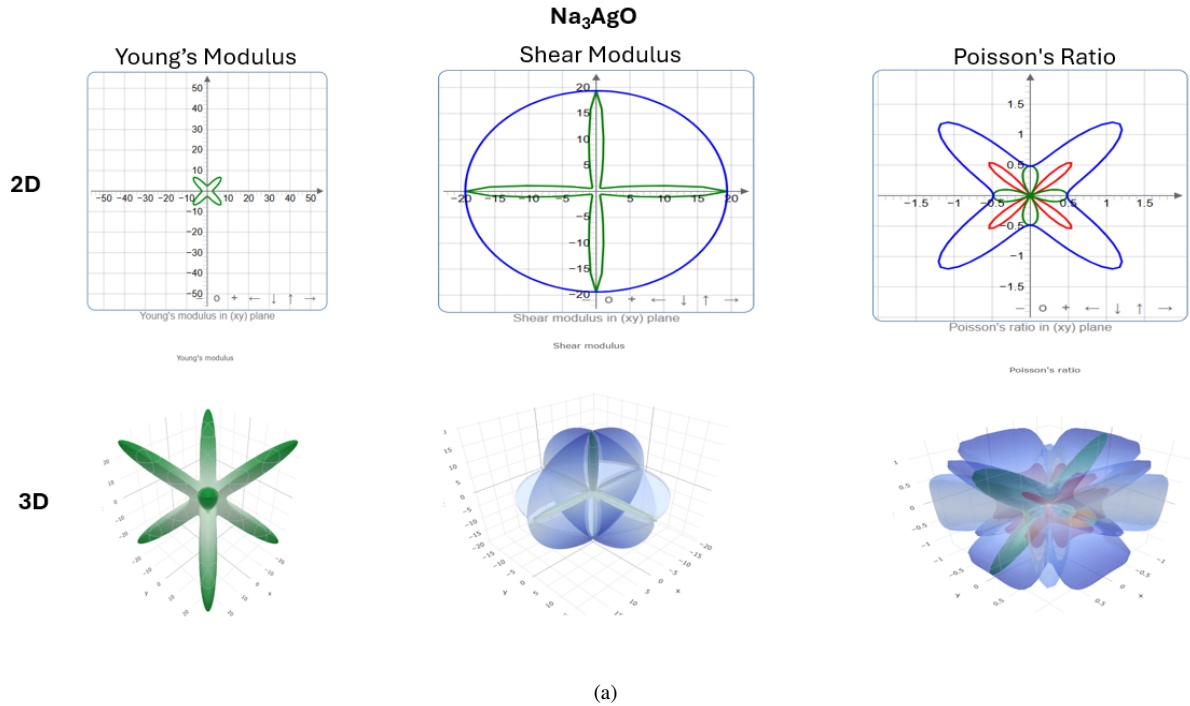

(a)

**Figure 4:** 2D and 3D graphs of elastic moduli for IP Na$_3$AgO

**Table 3**
Computed maximal and minimal ranging values of elastic moduli along with their anisotropic parameter for IP Na$_3$AgO

| Parameters | | Na$_3$AgO |
|---|---|---|
| Y(GPa) | Y$_{min}$ | 2.6902 |
| | Y$_{max}$ | 48.083 |
| | A | 17.87 |
| G(GPa) | G$_{min}$ | 0.9055 |
| | G$_{max}$ | 19.375 |
| | A | 21.4 |
| $\nu$ | $\nu_{min}$ | -0.76221 |
| | $\nu_{max}$ | 1.7153 |
| | A | $\infty$ |

## 6. Optical Properties

For the production of solar cells and optoelectronic devices, a detailed analysis of the optical properties of materials is required. The optical properties of a material are described by its complex dielectric function, which is expressed as $\epsilon(\omega) = \epsilon_1(\omega) + \iota\epsilon_2(\omega)$. This function describes how the material reacts to external electromagnetic (EM) radiation [33]. The capacity of a material to become polarized by induced electric dipoles upon exposure to external electromagnetic radiation is the primary measure of the real dielectric constant, $\epsilon_1(\omega)$. The imaginary part, $\epsilon_2(\omega)$, represents the material's capacity to transmit and attenuate electromagnetic radiation and are given as below [34, 35]

$$\epsilon_1(\omega) = 1 + \frac{2}{\pi}P\int_0^\infty \frac{\omega' \epsilon_2(\omega')}{\omega'^2 - \omega^2}d\omega' \quad (15)$$

$$\epsilon_2(\omega) = -\frac{2\omega}{\pi}P\int_0^\infty \frac{\epsilon_1(\omega)}{\omega'^2 - \omega^2}d\omega' \quad (16)$$

The photon frequency, represented by the symbol $\omega$ in this instance, determines the energy of the electromagnetic radiation interacting with the substance. By taking into account its Cauchy principal value and removing singularities, the word P, which denotes the integral's primary value, guarantees an appropriate evaluation of the integral. The $\epsilon_1(\omega)$ for our material is shown in fig. 5a. When the material exposed to electromagnetic radiation, the values of real dielectric function $\epsilon_1(\omega)$ show the level of polarization. As the amount of photon energy start increases, the initial peaks of the $\epsilon_1(\omega)$ value for Na$_3$AgO appear roughly at 11.57 for 0.71 eV . At higher energy levels, a sharp decrease is seen after these peaks. For Na$_3$AgO , the static amounts for $\epsilon_1(0)$ occur at 9.59. Higher $\epsilon_1(0)$ are beneficial for effective charge transport and improving the performance of solar cells [36]. The effectiveness of optoelectronic devices can be increased by using a material with a greater $\epsilon_1(0)$ because it has a reduced rate of charge recombination. The optical shifts between likely degrees of energy are primarily responsible for the $\epsilon_1(\omega)$ peaks in the 0.0–2.5 eV interval. Changes in the energy gap cause the $\epsilon_1(\omega)$ value to increase. The material under study have positive and negative $\epsilon_1(\omega)$ values over the whole range of computed energies, as shown in fig. 5a, indicating both light reflectance and absorption between 0.0 and 14 ev. Fig. 5b. shows the $\epsilon_2(\omega)$ as a function of photon energy, illustrates the absorption behavior. Na$_3$AgO show highest $\epsilon_2(\omega)$ peaks at 13.06 at 2.30 eV. Such type of transitions occur within the Brillouin zone (BZ) with respect to certain symmetrical directions (R, Γ, X, M, and Γ), as seen in fig. 2. The larger values of both the $\epsilon_1(\omega)$ and $\epsilon_2(\omega)$ at a low energy level along with smaller values of both the $\epsilon_1(\omega)$ and $\epsilon_2(\omega)$ at higher energy regions show the potential of this compound in microelectronics as well as integrated circuits. Also It is further noted that the real part of dielectric function becomes negative in the energy ranges (2.76 – 6.20) eV as materials behave metallic for negative values of $\epsilon_1(\omega)$ and is dielectric otherwise. The refractive index (fig. 5c.), or n($\omega$), is very crucial property for manufacturing a solar panel. it is given by the following equation as below [37]

$$n(\omega) = \sqrt{\frac{\epsilon_1(\omega)}{2} + \frac{\sqrt{\epsilon_1^2(\omega) + \epsilon_2^2(\omega)}}{2}} \quad (17)$$

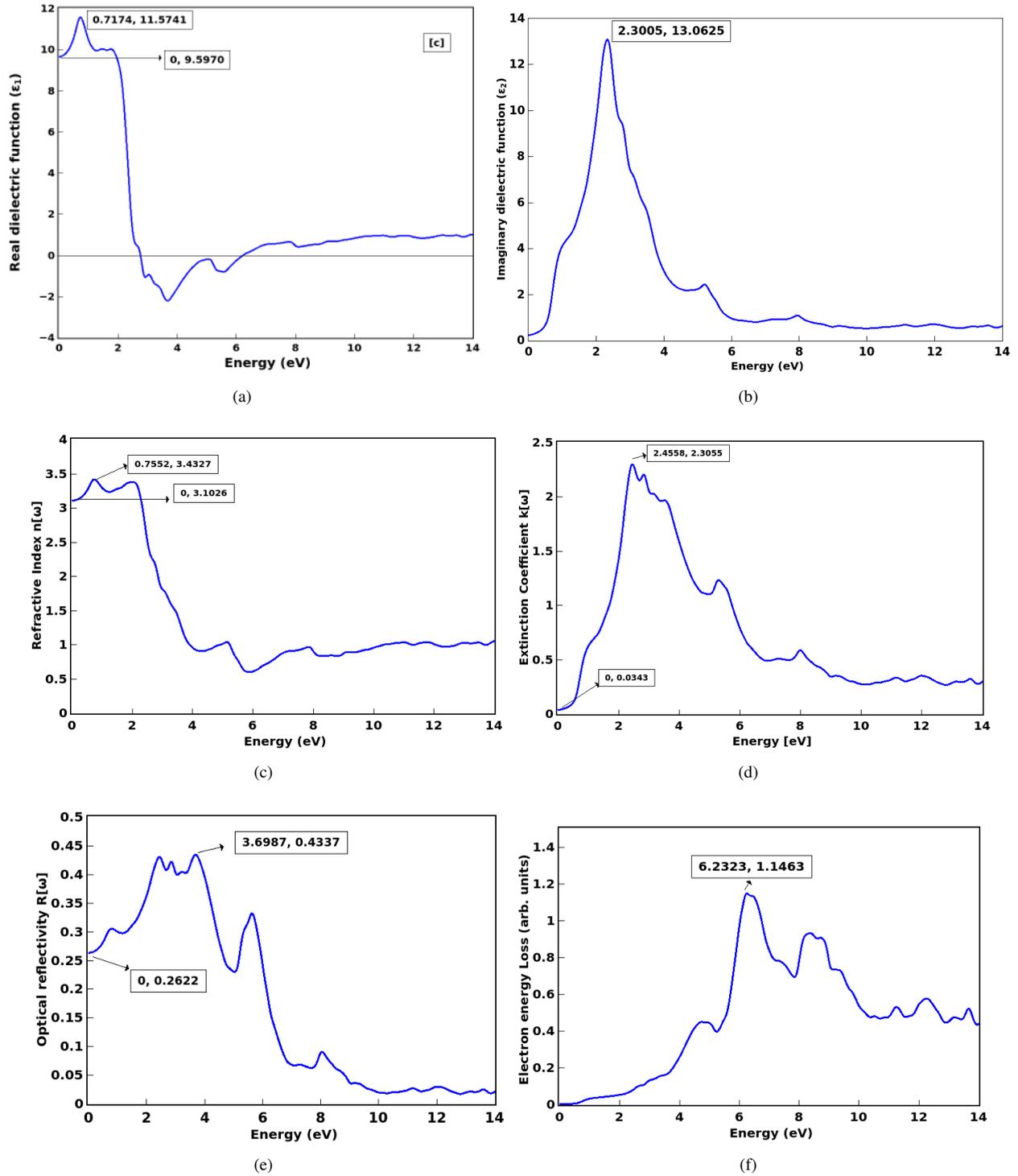

**Figure 5:** (a)–(f) Variation of real and imaginary parts of the dielectric function, Refractive index, Extinction coefficient, Optical reflectivity and Electron energy loss as a function of photon energy for the IP Na$_3$AgO

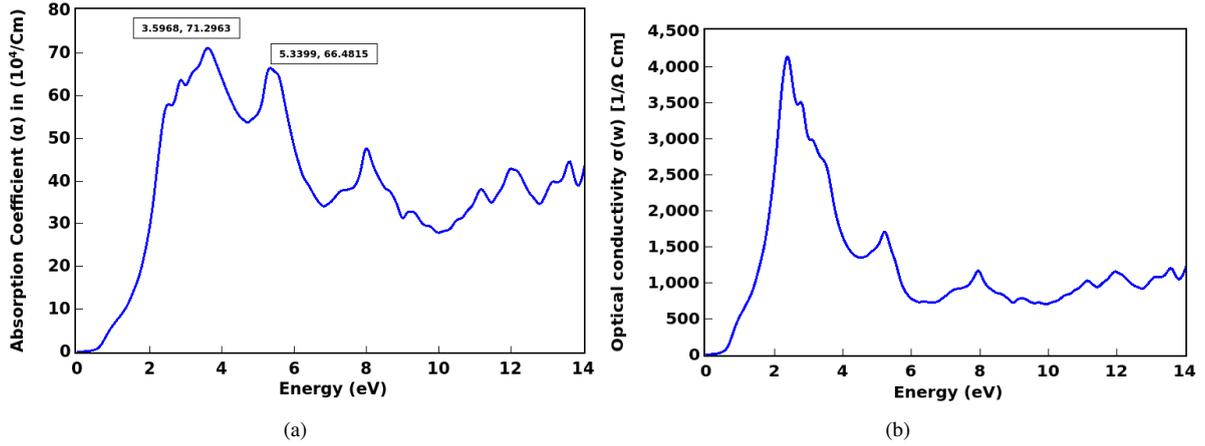

**Figure 6:** (a)–(b) Variation of Absorption coefficient and Optical conductivity as a function of photon energy for the IP Na$_3$AgO

The greatest refractive index numbers for Na$_3$AgO is 3.43 and at the zero frequency n(0), the calculated refracted index is 3.10. The larger n($\omega$) values at the mentioned energies point to possible uses for these anti-perovskites as transparent materials. Similar to $\epsilon_2(\omega)$, the extinction coefficient K($\omega$) measures the amount of energy absorbed by the substance. it is given by the following equation as given below

$$K(\omega) = \sqrt{\frac{-\epsilon_1(\omega)}{2} + \frac{\sqrt{\epsilon_1^2(\omega) + \epsilon_2^2(\omega)}}{2}} \quad (18)$$

From the fig. 5d., the graph between extinction coefficient K($\omega$) vs photon energy, it is clear that the highest attenuation of the wave due to both absorption and scattering occur at 2.45 at 2.30 eV, and at zero frequency it is about 0.034. The movement of carriers leads to the conduction or transportation of carriers after the photons are absorbed and attenuated by the substance [38]. Furthermore, the graphical representation of K($\omega$) shows resemblance to $\epsilon_2(\omega)$, indicating that these anti-perovskites attenuate more visible and UV light, ensuring higher charge transport in this region. One important optical property that describes how a material's surface responds to electromagnetic radiation is reflectivity R($\omega$) [39]. It is a crucial standard for characterizing materials and evaluating their potential applications. R($\omega$) is a measure of the ratio of incident to reflected light intensity on a material's surface. It is given by the following expression as below

$$R = \frac{(n-1)^2 + k^2}{(n+1)^2 + k^2} \quad (19)$$

The plot of reflection parameter R($\omega$), the important parameter in understanding the optical properties of a material, vs photon energy is shown in fig. 5e. Reflectivity achieve its maximum in energy range (0.0 – 4.2) eV. The maximum value of R($\omega$) in visible region of electromagnetic spectrum is 0.43 at 3.69 eV when real part of dielectric is negative, whereas static reflectivity is at 0.26. Further with the increase of photon energy reflectivity sharply decreases. The electron energy loss factor L($\omega$) reveals the optical energy loss caused by electron heating or scattering as electrons pass through a material [40]. It is expressed as

$$L(\omega) = Im\left[\frac{-1}{\epsilon}\right] \quad (20)$$

The electron energy loss graph, as shown in fig. 5f., show the maximum peak occur at 1.14(6.23 eV) suggesting specific excitation occur after energy of photon lost during the interaction with matter. Broader peaks or area under curve also suggest stronger energy loss or transferred within a specific range. Materials with minimal L($\omega$) are considered appropriate for prospective solar cell applications. The quantity of light at a particular wavelength that a substance absorbs prior to the absorption coefficient measuring transmission ($\alpha(\omega)$ [41], which is determined by Equation given as,

$$\alpha(\omega) = \sqrt{2}\omega\left[\sqrt{\epsilon_1(\omega)^2 + \epsilon_2(\omega)^2} - \epsilon_1(\omega)\right] \quad (21)$$

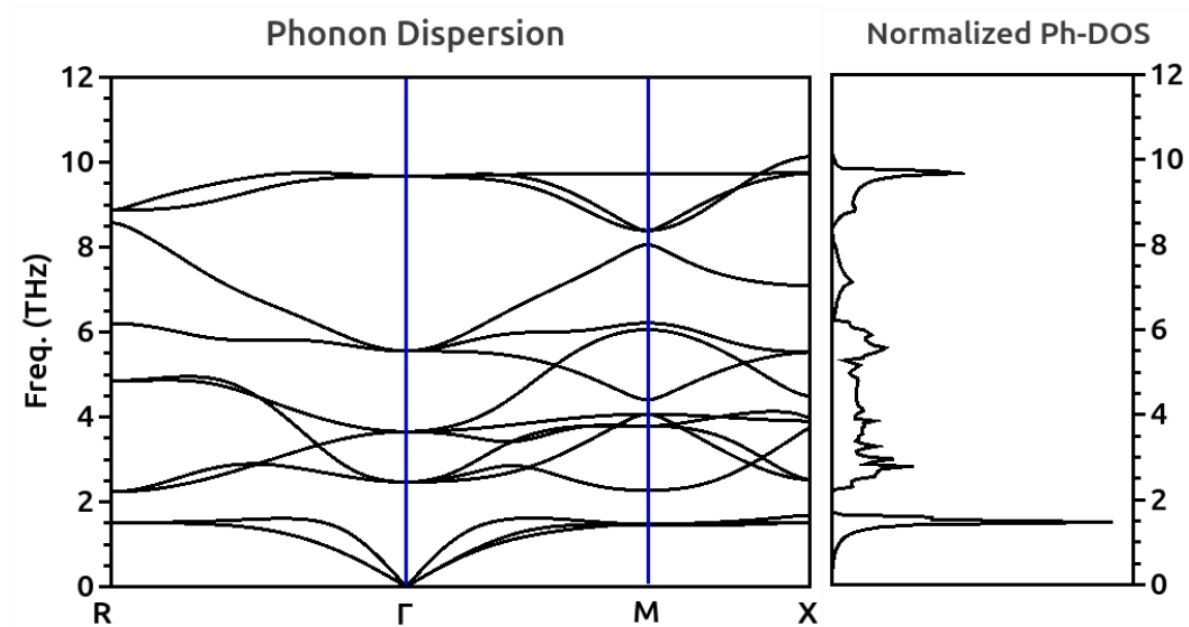

(a)

**Figure 7:** Phonon spectra obtained for IP Na$_3$AgO

The compound under study show its transparency to infrared region such that the absorption coefficient $\alpha(\omega)$ has approximately zero value below the 1.7 eV. In the spectrum of light we observed the greatest peak 71.29 (104/cm) occur at 3.59 eV which fall in ultra violet region that indicate the material under study could be used as absorbing material in solar cell. Finally, equation given as below

$$\sigma(\omega) = Re\left[-i\frac{\omega}{4\pi}(\epsilon(\omega)-1)\right] \qquad (22)$$

determines the optical conductivity $\sigma(\omega)$, which controls the conduction or transmission of charge carriers after the absorbed radiations. The graph between optical conductivity $\sigma(\omega)$ and photon energy shown in fig. 6b.. Material response to the oscillating electric field is found highest at 4125/Ωcm at 2.35 eV. The peaks of the graph correspond to electronics transition from valance band to conduction band. The computed result show that the loss function and reflectance have a close relationship that is in opposite direction, meaning that the loss function decreases when reflectivity peaks and vice versa. This development is particularly apparent in the visible and ultra-violet region where visible light is absorbed. According to these results, Na$_3$AgO is suitable anti-perovskites for use in optoelectronic devices.

## 7. Dynamic Stability

Quantized lattice vibrations within the crystal lattice are referred to as phonons, and phonon dispersion (PD) spectra are crucial for analyzing and verifying the dynamic stability. In the present work PD has been computed using Phonopy package with displacement method [42]. In energy conversion technologies, materials with dynamic stability may offer improved long-term dependability and durability. As shown in fig. 7 for Na$_3$AgO, the PD spectra are plots between the wavevector and vibrational mode frequency along symmetric spots in the Brillouin zone. It is crucial to remember that PD curves with frequencies above zero (positive) are thought to be stable, but dynamic instability is ensured by the presence of frequencies below zero (negative) [43]. Since Na$_3$AgO have frequencies of lattice vibrations above zero (positive), these IPs are dynamically stable.

## 8. Thermodynamic Properties

By looking at the thermodynamic response, we can determine the system's state and how it responds to outside energy and work. This study examines the alloy's specific heat, thermal expansion coefficient, Debye temperature, and Grüneisen parameter in response to changes in temperature and pressure, among other thermodynamic properties. First-principles density functional theory (DFT) calculations combined with the quasi-harmonic approximation (QHA) allow us to determine free energy, entropy, heat capacity, and related parameters across a variety of temperatures. We calculated the thermodynamic characteristics of the IP Na$_3$AgO at a pressure of 0 GPa and a temperature range of 0 to 400 K using the Quasiharmonic approximation and the GIBBS2 method [44]. These calculations are necessary to comprehend the severe conditions needed for thermodynamic applications. The Debye temperature ($\theta_D$) is the temperature at which the vibrations of a material attain their highest frequencies. The Debye temperature can be found

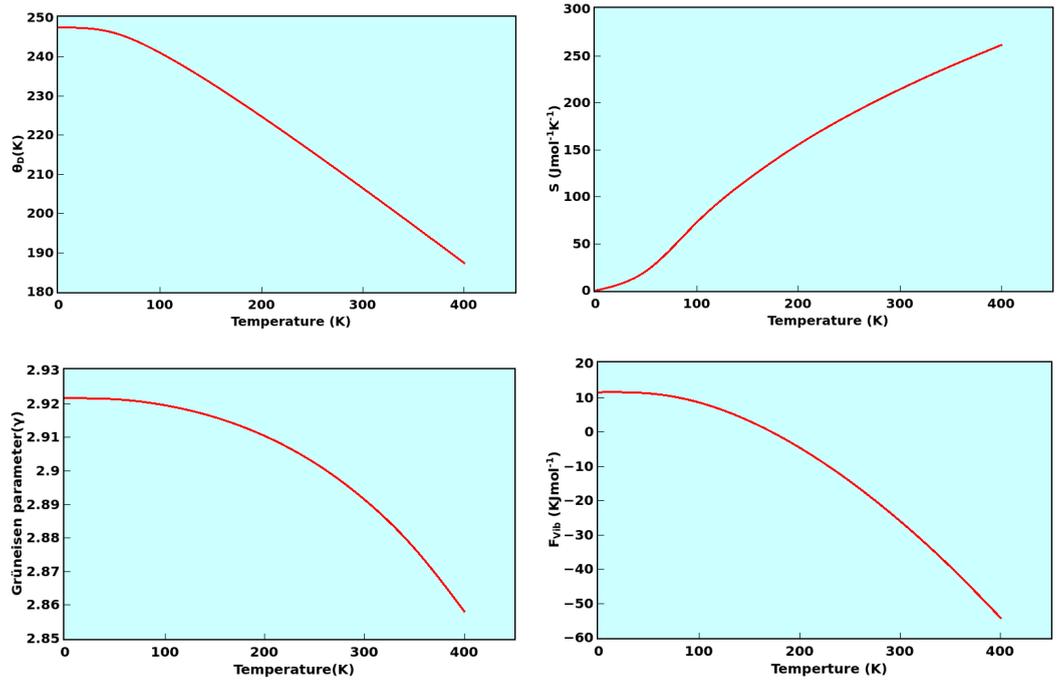

**Figure 8**: Variation of [a] Debye temperature [b] Entropy [c] Gruneisen parameter [d] Vibrational free energy $F_{vib}$ with temperature for IP Na$_3$AgO

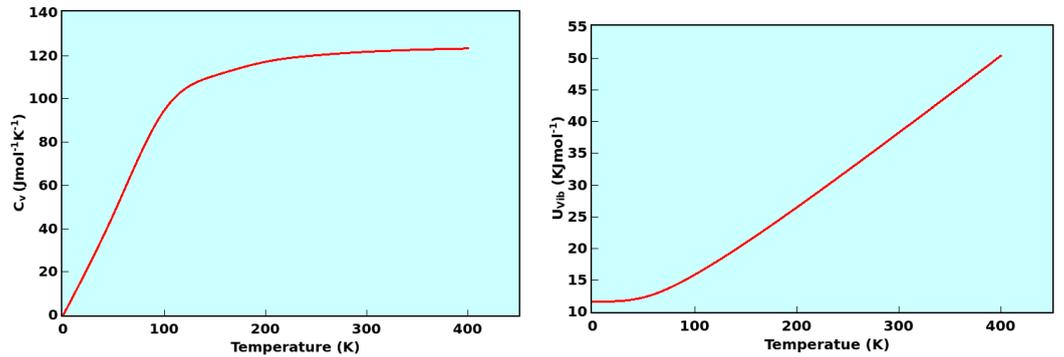

**Figure 9**: Variation of [a] Cv [b] Vibrational internal energy $U_{vib}$ with temperature for IP Na$_3$AgO

using the following formula:

$$\theta_D = \frac{h}{k_B}\left[\frac{3n}{4\pi}(\frac{\rho N_A}{M})\right]^{\frac{1}{3}} V_M \quad (23)$$

$$V_M = \frac{1}{3}\left[\frac{2}{V_t^3} + \frac{1}{V_L^3}\right] \quad (24)$$

$$V_L = \sqrt{\frac{3B + 4G}{3\rho}} \quad (25)$$

$$V_t = \sqrt{\frac{G}{\rho}} \quad (26)$$

Equations (23-26) can be used to determine particular thermodynamic parameters, such as longitudinal velocity, transverse velocity, and mean velocity of lattice vibrations, using the shear modulus (G) and bulk modulus (B). As per eqn. 23, the temperature dependence of IP Na$_3$AgO has been shown in fig.8a. It has been noted that the Debye temperature value drops exponentially with a notable increase in temperature. One important thermodynamic parameter that is mostly impacted by bonding strength, atomic mass distribution, and lattice dynamics is the Debye temperature. It gauges the crystal lattice's greatest vibrational frequency and is closely correlated with the material's sound speed and chemical bond stiffness. Figure 8a illustrates how the Debye temperature noticeably drops as the temperature rises. This effect happens when atomic vibrations intensify with increasing temperature, resulting in anharmonicity and a decrease in phonon frequencies. Furthermore, the strong covalent-metallic interaction between the sodium (Na), arsenic (As), and oxygen (O) atoms in the Na3AgO IP is responsible for the change in the Debye temperature. Fig. 9b. illustrates how the vibrational free energy of the IP

Na$_3$AgO rises with temperature due to greater lattice vibrations and a corresponding rise in phonon population at higher temperatures. The atoms of the crystal lattice oscillate more forcefully when heat energy is given, increasing the vibrational entropy and lowers the Helmholtz free energy. Thermodynamic stability is characterized by the alloy's intrinsic phonon softening and anharmonicity. The Grüneisen parameter (Γ), which gauges anharmonicity, is taken into consideration in order to comprehend lattice thermal conductivity. Controlled phonon scattering is indicated by moderate values of the Grüneisen parameter, which is essential for assessing thermoelectric efficiency. Only low-energy acoustic phonon modes contribute to vibrational free energy (Fvib) at low temperatures, as illustrated in fig. 9d., which causes a progressive decline. Fvib decreases more quickly as the temperature rises because higher-energy photonic modes are activated. By providing negative free energy to balance the internal energy and reduce the alloy's total Gibbs free energy, this temperature-dependent activity stabilizes the system. The dynamical stability of the system is confirmed by the lack of imaginary phonon frequencies across the Brillouin zone, suggesting that the vibrational contributions are physically significant and not the consequence of lattice instabilities. Furthermore, IP Na$_3$AgO has very soft vibrational phonon modes in alkali atoms, which results in larger vibrational entropy contributions that improve thermal stability at high temperatures. The smooth and continuous character of the vibrational free energy curve indicates the lack of phase transitions or structural distortions throughout the temperature range under investigation (up to about 400 K). Heat is another crucial thermodynamic parameter to take into account. Up to 300 K, the heat capacity (C$_v$) is proportional to T$^3$, beyond which it approaches saturation and obeys the Dulong-Petit rule, as shown in fig. 10a. Figure 10b illustrates how increased phonon activity, which comes from greater atomic vibrations brought on by thermal stimulation, causes the vibrational internal energy to rise with temperature. The enhanced thermal stimulation of lattice vibrations (phonons) in IP Na$_3$AgO causes the vibrational internal energy to increase with temperature. Atoms fluctuate more amplitudearily around their equilibrium locations as the temperature rises, adding to vibrational energy and filling additional phonon modes. The Bose-Einstein distribution and the phonon density of states, which both rise with temperature in accordance with the harmonic approximation in density functional theory (DFT), explain this phenomenon. More atomic vibrations and anharmonic effects within the crystal lattice of NaVAs Heusler alloys are indicated by the higher temperature results. Atoms fluctuate more amplitudearily around their equilibrium locations as the temperature rises, adding to vibrational energy and filling additional phonon modes. The Bose-Einstein distribution and the phonon density of states, which both rise with temperature in accordance with the harmonic approximation in density functional theory (DFT), explain this phenomenon. More atomic vibrations and anharmonic effects within the crystal lattice of NaVAs Heusler alloys are indicated by the higher temperature results. Thus, Na$_3$AgO, an oxide based on alkali metals, has a cubic crystal structure and exhibits exceptional thermodynamic behavior, indicating its stability and promise for high-temperature applications.

## 9. Conclusion

The main focus of this study was the attractive properties of inverse perovskite oxides Na$_3$AgO photoelectric conversion capability. The cubic structure and space group of pm-3 m were demonstrated by the structural features of IPs Na$_3$AgO (221). The experimental formulation and thermal stability of IPs were confirmed by the formation energy which is -2.27 eV/atom. The investigated IPs' dynamic stability and bonding strength against distortion of the cubic structure were confirmed by the analysis of phonon dispersion curves, which showed no negative frequencies. The energy gap (E$_g$) was estimated using the electronic properties, features the semiconducting behavior of IPs. Furthermore, Na$_3$AgO showed an indirect (E$_g$) of 1.273 eV. The feasibility of using Na$_3$AgO for solar cells was also confirmed by the identified light-dependent properties. An absorption value of 71.29 (10$^4$ cm$^{-1}$) was obtained, which is dependable for solar cells. Furthermore, ECs verified the mechanical stability, and additional elastic characteristics predicted the ductility, strength, hardness, and anisotropy of Na$_3$AgO. In the visible spectrum, Na$_3$AgO exhibits considerable absorption, scattering losses, and a notably low reflectivity. Consequently, the results of the analysis clearly imply that these fascinating anti-perovskites are potentially useful substances for solar energy harvesting or photovoltaic applications.

## Declaration of Competing Interest

The authors declare that they have no known competing financial interests or personal relationships that could have appeared to influence the work reported in this paper.

## Acknowledgement

Authors gratefully acknowledge the support by Sri Sai University Palampur and Department of Higher Education, Government of Himachal Pradesh.

## CRediT authorship contribution statement

**Vipan Kumar:** Data compilation, Writing - original draft Writing - review and editing. **Shyam Lal Gupta:** Conceptualization, Methodology, Data curation, Writing - original draft Writing - review and editing. **Sumit Kumar:** Software, Workstation, Data generation. **Ashwani Kumar:** Software, Workstation, Data generation. **Pooja Rana:** Data compilation, Writing - original draft Writing - review and editing. **Diwaker:** Conceptualization, Methodology, Data curation, Writing - original draft Writing - review and editing.